\documentclass[prb,twocolumn,byrevtex,floatfix,superscriptaddress]{revtex4}

\usepackage{amssymb,amsmath,amsfonts,revsymb,graphicx,afterpage}
\usepackage{color,bm}
\pdfoutput=1

\def \ni {{n_i}}

\def \vH {{\bf H}}

\def \vS {{\bf S}}

\def \vQ {{\bf Q}}

\def \mb {\mu_{\rm B}}

\def \zp {{\bf z}^{\prime }}

\def \vP {{\bf P}}

\def \BF {{\rm BiFeO$_3$} }

\def\wn/{\,cm$^{-1}$}
\def\area/{\,cm$^{-2}$}
\def\BFO/{BiFeO${_3}$}
\def\cubic/{$_\mathrm{c}$}
\def\DM/{Dzyaloshinskii-Moriya}

\def\bcgo/{Ba$_2$CoGe$_2$O$_7$}
\def\ccso/{Ca$_2$CoSi$_2$O$_7$}
\def\nfbo/{NdFe$_3$(BO$_3$)$_4$}
\def\tfbo/{TbFe$_3$(BO$_3$)$_4$}
\def\bfo/{BiFeO$_3$}
\def\azurite/{Cu$_3$(CO$_3$)$_2$(OH)$_2$}
\def\lco/{LiCu$_2$O$_2$}
\def\tmo/{TbMnO$_3$}

\def\edc/{\ensuremath{\mathbf{E}_0}}
\def\eac/{\ensuremath{\mathbf{E}_1}}
\def\bdc/{\ensuremath{\mathbf{B}_0}}
\def\bac/{\ensuremath{\mathbf{B}_1}}

\begin{document}

\title{\boldmath Optical diode effect in the room-temperature
multiferroic BiFeO$_3$ \unboldmath}

\author{I. K\'ezsm\'arki}
\affiliation{ Department of Physics, Budapest University of
Technology and Economics and MTA-BME Lend\"ulet Magneto-optical
Spectroscopy Research Group, 1111 Budapest, Hungary}

\author{U. Nagel}
\affiliation{National Institute of Chemical Physics and Biophysics,
12618 Tallinn, Estonia}

\author{S. Bord\'acs}
\affiliation{ Department of Physics, Budapest University of
Technology and Economics and MTA-BME Lend\"ulet Magneto-optical
Spectroscopy Research Group, 1111 Budapest, Hungary}

\author{R. S. Fishman}
\affiliation{Materials Science and Technology Division, Oak Ridge
National Laboratory, Oak Ridge, Tennessee 37831, USA}

\author{J. H. Lee}
\affiliation{Materials Science and Technology Division, Oak Ridge
National Laboratory, Oak Ridge, Tennessee 37831, USA}

\author{H. T. Yi}
\affiliation{Rutgers Center for Emergent Materials and Department of
Physics and Astronomy, Rutgers University, Piscataway, New Jersey
08854, USA}

\author{S.-W. Cheong}
\affiliation{Rutgers Center for Emergent Materials and Department of
Physics and Astronomy, Rutgers University, Piscataway, New Jersey
08854, USA}

\author{T. R{\~o}{\~o}m}
\affiliation{National Institute of Chemical Physics and Biophysics,
12618 Tallinn, Estonia}

\maketitle

\textbf{Multiferroics permit the magnetic control of the electric
polarization and electric control of the
magnetization~\cite{Fiebig2005,Eerenstein2006,Kimura2007,Cheong2007,Martin2010,Sando2013,Henron2014,Matsukura2015}.
These static magnetoelectric (ME) effects are of enormous interest:
The ability to read and write a magnetic state current-free by an
electric voltage would provide a huge technological
advantage~\cite{Martin2010,Sando2013,Henron2014,Matsukura2015}.
Dynamic or optical ME effects are equally interesting because they
give rise to unidirectional light propagation as recently observed
in low-temperature
multiferroics~\cite{Arima2008,Kezsmarki2011,Bordacs2012,Takahashi2012,Takahashi2013,Kezsmarki2014}.
This phenomenon, if realized at room temperature, would allow the
development of optical diodes which transmit unpolarized light in
one, but not in the opposite direction. Here, we report strong
unidirectional transmission in the room-temperature multiferroic
BiFeO$_3$ over the gigahertz--terahertz frequency range. Supporting
theory attributes the observed unidirectional transmission to the
spin-current driven dynamic ME effect. These findings are an
important step toward the realization of optical diodes,
supplemented by the ability to switch the transmission direction
with a magnetic or electric field.}

BiFeO$_3$ is by the far most studied compound among multiferroic and
magnetoelectric materials. While experimental studies have already
reported about the first realizations of the ME memory function
using BiFeO$_3$ based
devices~\cite{Martin2010,Sando2013,Henron2014,Matsukura2015}, the
origin of the ME effect is still under debate due to the complexity
of the material. Because of the low symmetry of iron sites and
iron-iron bonds, the magnetic ordering can induce local polarization
via each of the three canonical terms~\cite{Jia} -- the
spin-current, exchange-striction and single-ion mechanisms. While
the spin-current term has been identified as the leading
contribution to the magnetically induced ferroelectric polarization
in various studies~\cite{katsura05,Cheong2007,Tokunaga2015}, the
spin-driven atomic displacements~\cite{Lee2013} and the electrically
induced shift of the spin-wave (magnon)
resonances~\cite{Rovillain2010} were interpreted based on the
exchange-striction and single-ion mechanisms, respectively.
%Due to the extremely long (62\,nm) pitch of the spin
%cycloids~\cite{Sosnowska1982} and the antiferromagnetic spin
%ordering between neighbouring cycloids in BiFeO$_3$, the
%neighbouring spins are nearly parallel or antiparallel to each other
%spanning an angle less than 2 degrees. Since  the exchange striction
%mechanism can efficiently generate polarization in collinear spin
%structures unlike the spin-current mechanism, it may even play a
%more important role in the static ME properties of BiFeO$_3$. In
%particular, the exchange-striction and single-ion mechanisms were
%found indispensable to respectively reproduce the spin-driven atomic
%displacements~\cite{Lee2013} and the electrically induced shift of
%the spin-wave (magnon) resonances~\cite{Rovillain2010}. Here we show
%that the dynamic ME effect at magnon resonances is dominated by the
%spin-current mechanism.

In the magnetically ordered phase below $T_N$$=$640\,K, BiFeO$_3$
possesses an exceptionally large spin-driven
polarization~\cite{Lee2013}, if not the largest among all known
multiferroic materials. Nevertheless, its systematic study has long
been hindered by the huge lattice ferroelectric polarization
($\mathbf{P}_0$) developing along one of the cubic
$\langle$111$\rangle$ directions at the Curie temperature
$T_C$$=$1100\,K and by the lack of single-domain ferroelectric
crystals. Owing to the coupling between $\mathbf{P}_0$ and the
spin-driven polarization, in zero magnetic field they both point
along the same [111] axis. A recent systematic study of the static
ME effect revealed additional spin-driven polarization orthogonal to
[111]~\cite{Tokunaga2015}.

\begin{figure}[t!]
\includegraphics[width=0.48\textwidth]{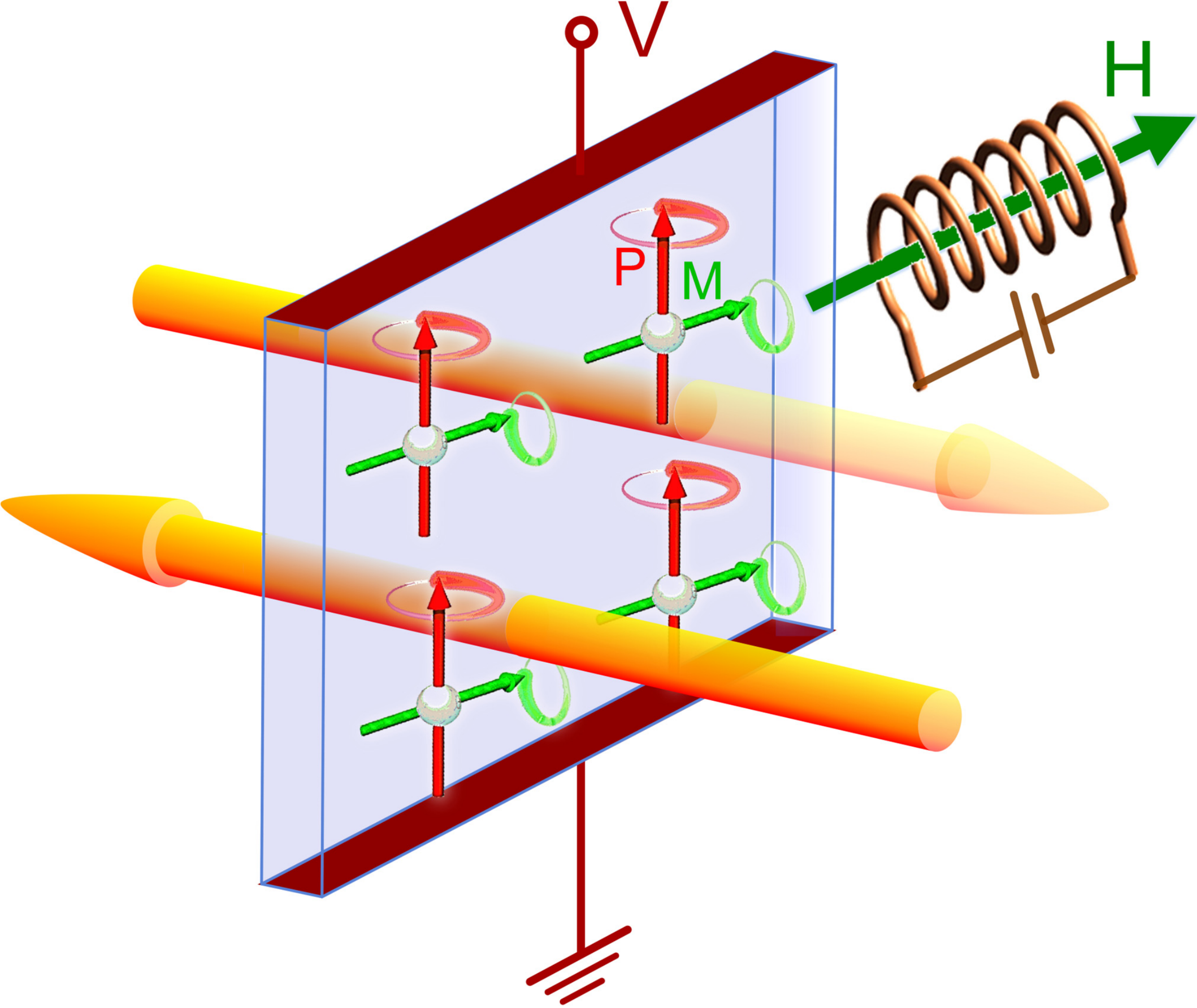}
\caption{\label{fig0} \textbf{$\mid$ Optical diode function in
multiferroics.} Ferro-type ordering of the local electric dipoles
(red arrows) and magnetic moments (green arrows) produces a
ferroelectric polarization $P$ and a spontaneous magnetization $M$,
respectively. Light interacts with both ferroic order parameters,
hence, upon illumination $P$ and $M$ oscillate coherently with the
electromagnetic field around their equilibria. The light induced
polarization has contributions from both the usual dielectric
permittivity and the optical ME effect $\chi^{em}(\omega)$. While
the first contribution is independent of the light propagation
direction, the polarization induced via the optical ME effect has
opposite sign for counter-propagating light beams. This can give
rise to either a constructive or a destructive interference between
the two terms. Similarly, the magnetization dynamics is governed by
the interference between the magnetization induced via the magnetic
permeability and the optical ME effect $\chi^{me}(\omega)$.
Consequently, the transmitted intensity depends on the propagation
direction (intense and pale yellow beams) even for unpolarized light
and can be exploited to produce optical diodes transmitting light in
one, but not in the opposite direction. The transmitting direction
can be reversed by switching the sign of either $P$ via an electric
voltage ($V$) or $M$ by an external magnetic field ($H$).}
\end{figure}

The optical ME effect of the magnon modes in multiferroics, which
gives rise to the unidirectional transmission in the
gigahertz--terahertz frequency range, has recently become a hot
topic in materials science. The difference in the absorption
coefficients ($\alpha$) of beams counterpropagating in such ME
media---called directional dichroism---can be expressed for linear
light polarization as~\cite{Kezsmarki2014,Miyahara2012}
\begin{equation}
\begin{split}
\Delta\alpha_{k}(\omega)=\alpha_{+k}(\omega)-\alpha_{-k}(\omega)\approx\frac{2\omega}{c}\Im\{\chi^{me}_{\gamma\delta}(\omega)-\chi^{em}_{\delta\gamma}(\omega)\}.
\label{dalpha_pm}
\end{split}
\end{equation}
The dynamic ME susceptibility tensors $\hat{\chi}^{me}(\omega)$ and
$\hat{\chi}^{em}(\omega)$ respectively describe the magnetization
generated by the oscillating electric field of light, $\Delta
M_{\gamma}^{\omega}=(\varepsilon_0/\mu_0)^{1/2}\chi^{me}_{\gamma\delta}(\omega)E_{\delta}^{\omega}$,
and the electric polarization induced by its oscillating magnetic
field, $\Delta
P_{\delta}^{\omega}=(\varepsilon_0\mu_0)^{1/2}\chi^{em}_{\delta\gamma}(\omega)H_{\gamma}^{\omega}$.
Here $\varepsilon_0$ and $\mu_0$ are the vacuum permittivity and
permeability, respectively, while  $\gamma$ and $\delta$ stand for
the Cartesian coordinates. Since the two cross-coupling tensors are
connected by the time-reversal operation $[\ldots]^{\prime}$
according to
$[\chi^{me}_{\gamma\delta}(\omega)]^{\prime}=-\chi^{em}_{\delta\gamma}(\omega)$,
the directional dichroism becomes
$\Delta\alpha_{k}(\omega)=\frac{2\omega}{c}\Im\{\chi^{me}_{\gamma\delta}(\omega)-[\chi^{me}_{\gamma\delta}(\omega)]^{\prime}\}$.
In other words, the directional dichroism emerges for simultaneously
electric- and magnetic-dipole active excitations and its magnitude
is determined by the time-reversal odd parts of the off-diagonal
$\chi^{me}_{\gamma\delta}(\omega)$ tensor
elements~\cite{Kezsmarki2011,Bordacs2012,Miyahara2012,Miyahara2011,Szaller2014}.
The schematic representation of the optical diode function in ME
media is shown in Fig.~\ref{fig0}.

\begin{figure}[t!]
% project file: model_T.pptx
%
\includegraphics[width=0.42\textwidth]{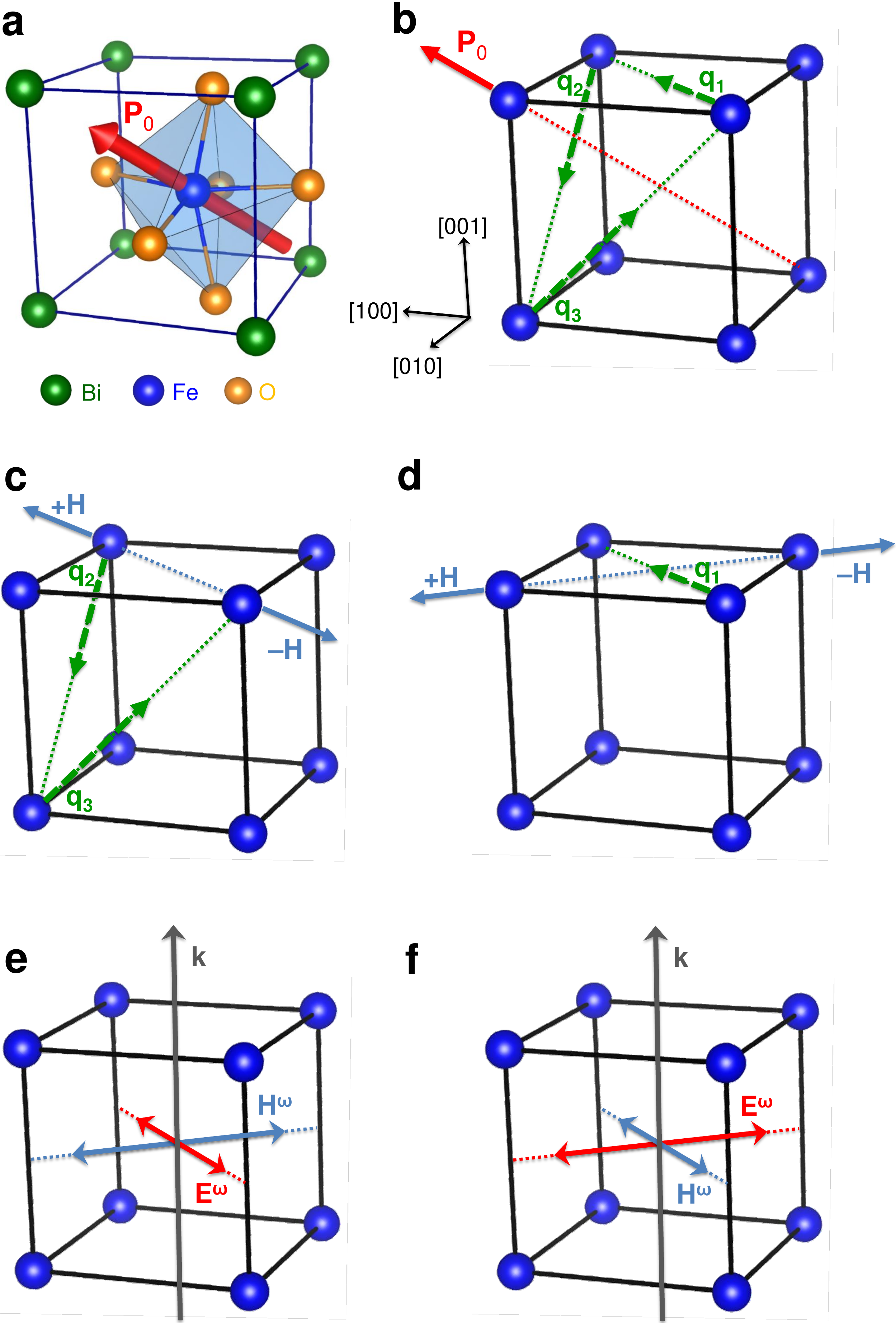}
\caption{\label{fig:1} \textbf{$\mid$ Experimental configurations
used to detect unidirectional transmission in BiFeO$_3$.}
\textbf{a,} Pseudocubic unit cell of BiFeO$_3$ showing the positions
of Bi, Fe and O ions. The lattice ferroelectric polarization,
$\mathbf{P}_0$$\parallel$[111], is schematically indicated on the Fe
site. \textbf{b,} Illustration of the three equivalent directions of
the cycloidal ordering vector $\mathbf{q}_i$ on the Fe sublattice.
The frame of reference is common to all panels. \textbf{c,} In
magnetic fields ($\pm H$) applied along [1$\bar{1}$0], cycloidal
domains with $\mathbf{q}_2$ and $\mathbf{q}_3$ are equally favoured,
while the domain with $\mathbf{q}_1$ is
suppressed~\cite{fishman13a,fishman13b}. \textbf{d,} In magnetic
fields ($\pm H$) applied along [110], only the cycloidal domain with
$\mathbf{q}_1$ is stable~\cite{fishman13a,fishman13b}. \textbf{e} \&
\textbf{f,} The propagation direction ($\mathbf{k}$) and the two
orthogonal polarizations of light beams traveling in the material.}
\end{figure}
%===========

In the cycloidal spin state of BiFeO$_3$~\cite{Sosnowska1982},
several low-frequency collective modes have been observed by
spectroscopic methods including light
absorption~\cite{nagel13,talbayev11} and Raman
spectroscopy~\cite{Cazayous2008,Rovillain2010,Sando2013}. Though the
electric-field-induced shift of the resonance frequencies observed
in the Raman study indicates the ME nature of these magnon
modes~\cite{Rovillain2010}, the optical ME effect has not been
investigated in BiFeO$_3$. Here, we performed absorption
measurements in the gigahertz--terahertz spectral range on
single-domain ferroelectric BiFeO$_3$ crystals~\cite{Talbayev2008}
with $\textbf{P}_0$ along [111] between room temperature and
$T$$=$4\,K in magnetic fields up to $\mu_0H$$=$17\,T. We found that
some of the magnon modes exhibit strong unidirectional transmission.
We identified the minimal set of spin-driven-polarization terms and
quantitatively reproduced both the spectral shape and the field
dependence of the directional dichroism solely by the spin-current
mechanism.

%BiFeO$_3$ samples used in the present study were confirmed to be
%polarization mono-domain single crystals with a ferroelectric
%polarization ($\mathbf{P}_0$) pointing along the [111]
%axis~\cite{Talbayev2008}.
The experimental configurations are schematically illustrated in
Fig.~\ref{fig:1}. Absorption spectra were obtained for light beams
propagating along [001] with two orthogonal linear polarizations,
$\mathbf{E}^{\omega}$$\parallel$[1$\overline{1}$0] and
$\mathbf{E}^{\omega}$$\parallel$[110]. Static magnetic fields ($\pm
H$) were applied perpendicular to the light propagation direction
along either [110] or [1$\overline{1}$0].

In simple magnets, such as ferromagnets, the sign change of the
magnetization corresponds to the time reversal operation. Thus, it
is equivalent to the reversal of the light propagation direction.
Owing to experimental limitations, in such cases, the absorption
change upon the magnetic field induced reversal of the
magnetization,
$\Delta\alpha_{H}$$=$$\alpha_{+H,+k}$$-$$\alpha_{-H,+k}$, is
typically detected instead of the absorption change associated with
the reversal of the light propagation direction,
$\Delta\alpha_{k}$$=$$\alpha_{+H,+k}$$-$$\alpha_{+H,-k}$. Though the
relation $\Delta\alpha_{k}=\Delta\alpha_{H}$ does not necessarily
hold for complex spin structures, such as BiFeO$_3$,
$\Delta\alpha_{k}$ and $\Delta\alpha_{H}$ spectra obtained from our
calculations are equal within 1-2\,\% for the experimental
configurations studied here.

%========================================================
\begin{figure*}[th!]
\includegraphics[width=0.75\textwidth]{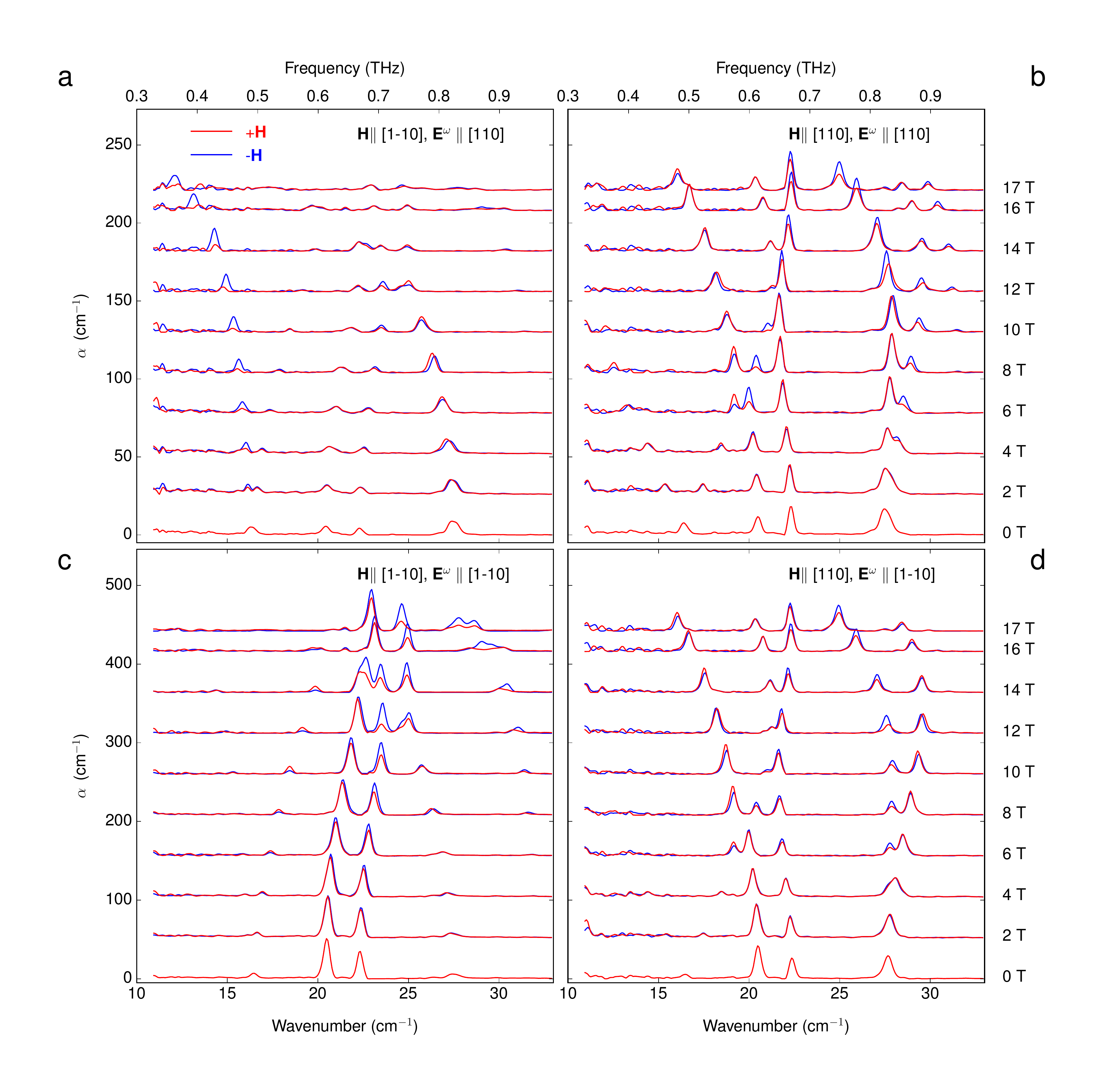}
\caption{\label{fig:Abs} \textbf{$\mid$ Absorption spectra of
BiFeO$_3$ in the range of magnon resonances.} \textbf{a-d} Magnetic
field dependent part of the absorption spectra measured at
$\mathrm{T}=2.5$\,K in four different configurations, i.e. for the
two orientations of the magnetic field ($\mathbf{H}$) and the two
orthogonal polarizations schematically shown in Fig.~\ref{fig:1}.
The light propagation direction is common to all experimental
configurations, $\mathbf{k}$$\parallel$[001]. Absorption spectra
measured in different magnetic fields are shifted vertically in
proportion to the magnitude of the field, and spectra recorded in
$+H$ and $-H$ are plotted with red and black lines, respectively.
Spectra shown in panel \textbf{a} \& \textbf{c} represent absorption
form the $\mathbf{q}_1$ cycloidal domain stabilized by
$\mathbf{H}$$\parallel$[1$\bar{1}$0], while spectra in panel
\textbf{b} \& \textbf{d} have contributions from $\mathbf{q}_2$ and
$\mathbf{q}_3$ domains favoured by $\mathbf{H}$$\parallel$[110].}
\end{figure*}
%===========

Figure~\ref{fig:Abs} shows the absorption spectra measured in four
different configurations, i.e. for two orientations of the magnetic
field and two light polarizations. The absorption coefficient at
several magnon resonances depends on the sign of the magnetic field.
This difference is stronger for
$\mathbf{H}$$\parallel$[1$\overline{1}0$] and most pronounced for
the lowest-frequency mode $\Psi_0$ in Fig.~\ref{fig:Abs}a when
$\mathbf{H}$$\parallel$[1$\overline{1}0$] and
$\mathbf{E}^{\omega}$$\parallel$[110]. With increasing magnetic
field this resonance becomes almost transparent for $+H$, while its
absorption increases for $-H$. We also measured the absorption
spectra with both light polarizations for
$\mathbf{H}$$\parallel$[001] and could not detect any difference
between $\pm H$.

In order to reproduce the observed directional dichroism on a
microscopic basis, we adopt the spin model of
Refs.[\onlinecite{fishman13a,fishman13b}], which successfully
describes the magnetic field dependence of the magnon resonances
(see the Methods section). Similarly to the static ME effect, all
the three basic mechanisms---the spin-current, exchange-striction
and single-ion mechanism---can in principle contribute to the
optical ME effect. By including all symmetry-allowed spin-driven
polarization terms, we calculated the optical ME susceptibilities
$\hat{\chi}^{me}(\omega)$ and $\hat{\chi}^{em}(\omega)$, the
dielectric permittivity $\hat{\varepsilon}(\omega)$ and the magnetic
permeability $\hat{\mu}(\omega)$~\cite{Kezsmarki2014,Miyahara2012}.
Next, we numerically solved the Maxwell equations by including these
response functions in the constitutive relations and calculated the
transmission of linearly polarized incoming beams for both backward
and forward propagation. The same calculation was done for both
field directions, $\pm H$. As already mentioned, we found
$\Delta\alpha_{k}\approx\Delta\alpha_{H}$ irrespective of the
magnitude of $H$.

To identify the spin-driven polarization terms relevant to the
optical ME effect, we performed a systematic fitting of the measured
$\Delta\alpha_{H}(\omega)$ by treating the magnitude of the
different terms as free parameters. We found that the directional
dichroism spectra are closely reproduced by the following two types
of spin-current terms
\begin{equation}
P_{\alpha}^{SC}=\frac{1}{N}\sum_{\langle
i,j\rangle}\{\lambda_{\alpha}^{(1)}[\mathbf{e}_{i,j}\times(\mathbf{S}_i\times\mathbf{S}_j)]_{\alpha}+(-1)^{n_i}\lambda_{\alpha}^{(2)}[\mathbf{S}_i\times\mathbf{S}_j]_{\alpha}\}\,
\end{equation}
where the summation goes over neighbouring spins connected by unit
vectors $\mathbf{e}_{i,j}$ and the integer $n_i$ labels the
hexagonal layers along [111]. The dynamic ME effect generated by the
spin-current terms is described by the coupling constants
$\lambda_{\alpha}^{(1)}$ and $\lambda_{\alpha}^{(2)}$, where
$\alpha$$=$$x^{\prime},y^{\prime},z^{\prime}$ stands for the three
coordinates along the axes
$\mathbf{x}^{\prime}$$\parallel$$\mathbf{q}_i$,
$\mathbf{y}^{\prime}$$\parallel$$(\mathbf{P}_0$$\times$$\mathbf{q}_i)$
and $\mathbf{z}^{\prime}$$\parallel$$\mathbf{P}_0$ (see
Fig.~\ref{fig:1}b).

\begin{figure}[t!]
\includegraphics[width=0.48\textwidth]{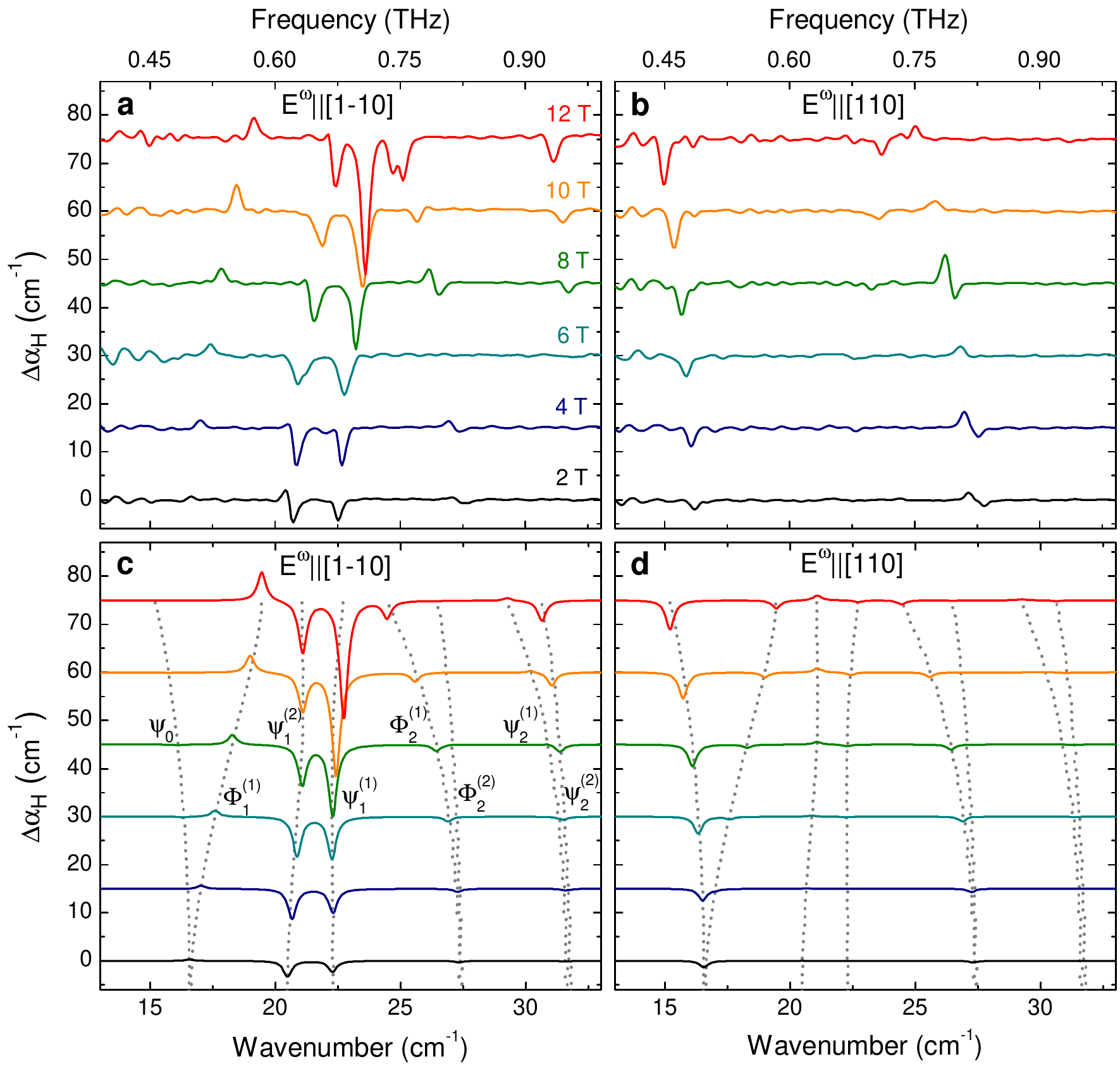}
\caption{\label{fig:theo-exp} \textbf{$\mid$ Directional dichroism
spectra of BiFeO$_3$ in the range of magnon resonances.} \textbf{a
\& b,} Magnetic field dependence of the directional dichroism
spectra measured as $\Delta\alpha_{H}(\omega)$ at $T$$=$2.5\,K with
the two orthogonal polarizations
$\mathbf{E}^{\omega}$$\parallel$[1$\overline{1}$0] and
$\mathbf{E}^{\omega}$$\parallel$[110], respectively. Spectra
obtained in different magnetic fields are shifted vertically in
proportion to the magnitude of the field, which was applied along
[1$\overline{1}$0]. The field values (common to each panel) are
indicated with labels on the top of the spectra in panel {\textbf
a}. \textbf{c} \& \textbf{d,} Directional dichroism spectra
predicted by our model for the case of panels \textbf{a} \&
\textbf{b}, respectively. The calculated mode frequencies are
indicated by dashed lines. For the assignment of the different modes
see Refs.[\onlinecite{fishman13a,fishman13b}].}
\end{figure}

Figure~\ref{fig:theo-exp} shows the comparison between the measured
and calculated directional dichroism spectra for
$\mathbf{H}$$\parallel$[1$\overline{1}0$] with the two orthogonal
light polarizations,
$\mathbf{E}^{\omega}$$\parallel$[1$\overline{1}$0] and
$\mathbf{E}^{\omega}$$\parallel$[110]. The best fit was obtained
with three independent parameters: $\lambda_{x^{\prime}}^{(1)}$$=$0,
$\lambda_{y^{\prime}}^{(1)}$$=$$-2\lambda_{z^{\prime}}^{(1)}$$\approx$57.0$\pm3.1$\,nC/cm$^{2}$,
$\lambda_{x^{\prime}}^{(2)}$$=$$\lambda_{y^{\prime}}^{(2)}$$\approx$34.5$\pm$2.4\,nC/cm$^{2}$,
$\lambda_{z^{\prime}}^{(2)}$$\approx$11.8$\pm2.9$\,nC/cm$^{2}$. The
population of the two cycloidal domains with $\mathbf{q_2}$ and
$\mathbf{q_3}$ propagation vectors was kept
equal~\cite{fishman13a,fishman13b}. We note that this limited set of
parameters provides only a semi-quantitative description of the mean
absorption spectra,
$\overline{\alpha}(\omega)$$=$$\alpha_{+H,+k}(\omega)$$+$$\alpha_{-H,+k}(\omega)$.

%========================================================
%===========
%A=B=-40.3\pm2.2
%gamma_x=gamma_y=-34.5\pm2.4
%gamma_z=-11.8\pm2.9
%lambda^(1)_{yp zp } = -2A
%lambda^(1)_{zp yp } = - B
%lambda^(2)_{xp } =\lambda^(2)_{yp } = gamma_x = gamma_y
%lambda^(2)_{zp } = gamma_z
We found that additional terms did not further improve the quality
of the fit. Hence, the optical ME effect in BiFeO$_3$ is dominated
by two types of spin-current polarizations, while the
exchange-striction and single-ion polarization terms do not
significantly contribute to it. This stems from the general nature
of the spin dynamics in BiFeO$_3$. Due to the very weak on-site
anisotropy acting on the $S=5/2$ iron spins, each magnon mode
corresponds to pure precessions of the spins, where the oscillating
component of the spin on site $i$, $\mathbf{\delta S}_i^{\omega}$,
is perpendicular to its equilibrium direction, $\mathbf{S}_i^0$.
This is in contrast to the spin stretching modes observed in highly
anisotropic magnets~\cite{Miyahara2011}. Since neighbouring spins
are nearly collinear in the cycloidal state with extremely long
(62\,nm) pitch~\cite{Sosnowska1982}, a dynamic polarization is
efficiently induced via spin-current terms such as $\mathbf{\delta
P}_i^{\omega}\propto \mathbf{S}_i^0\times\mathbf{\delta
S}_{i+1}^{\omega}$. In contrast, the dynamic polarization generated
by exchange-striction terms such as $\mathbf{\delta
P}_i^{\omega}\propto \mathbf{S}_i^0\cdot\mathbf{\delta
S}_{i+1}^{\omega}$ is nearly zero.

Despite its success in quantitatively describing the directional
dichroism spectra observed for
$\mathbf{H}$$\parallel$[1$\overline{1}0$], our model may not be
complete. When light propagates along [001], we predict that
directional dichroism should be absent for a magnetic field along
[$\eta\eta\kappa$]. While this is in agreement with
$\Delta\alpha_{H}$$=$$0$ found for $\mathbf{H}$$\parallel$[001], it
cannot account for the finite directional dichroism discerned in
Figs.~\ref{fig:Abs}c and \ref{fig:Abs}d for
$\mathbf{H}$$\parallel$[110]. This discrepancy may come from
additional anisotropy terms, neglected in the microscopic spin
Hamiltonian adopted from Refs.[\onlinecite{fishman13a,fishman13b}],
which further reduce the symmetry of the magnetic state and allow
the weak directional dichroism observed for
$\mathbf{H}$$\parallel$[110].

%===========
% project file:  C:\Users\Public\Documents\Samples\Multiferroics\BiFeO3\BiFeO3_Directional_effects.opj
\begin{figure}[t!]
\includegraphics[width=2.7in]{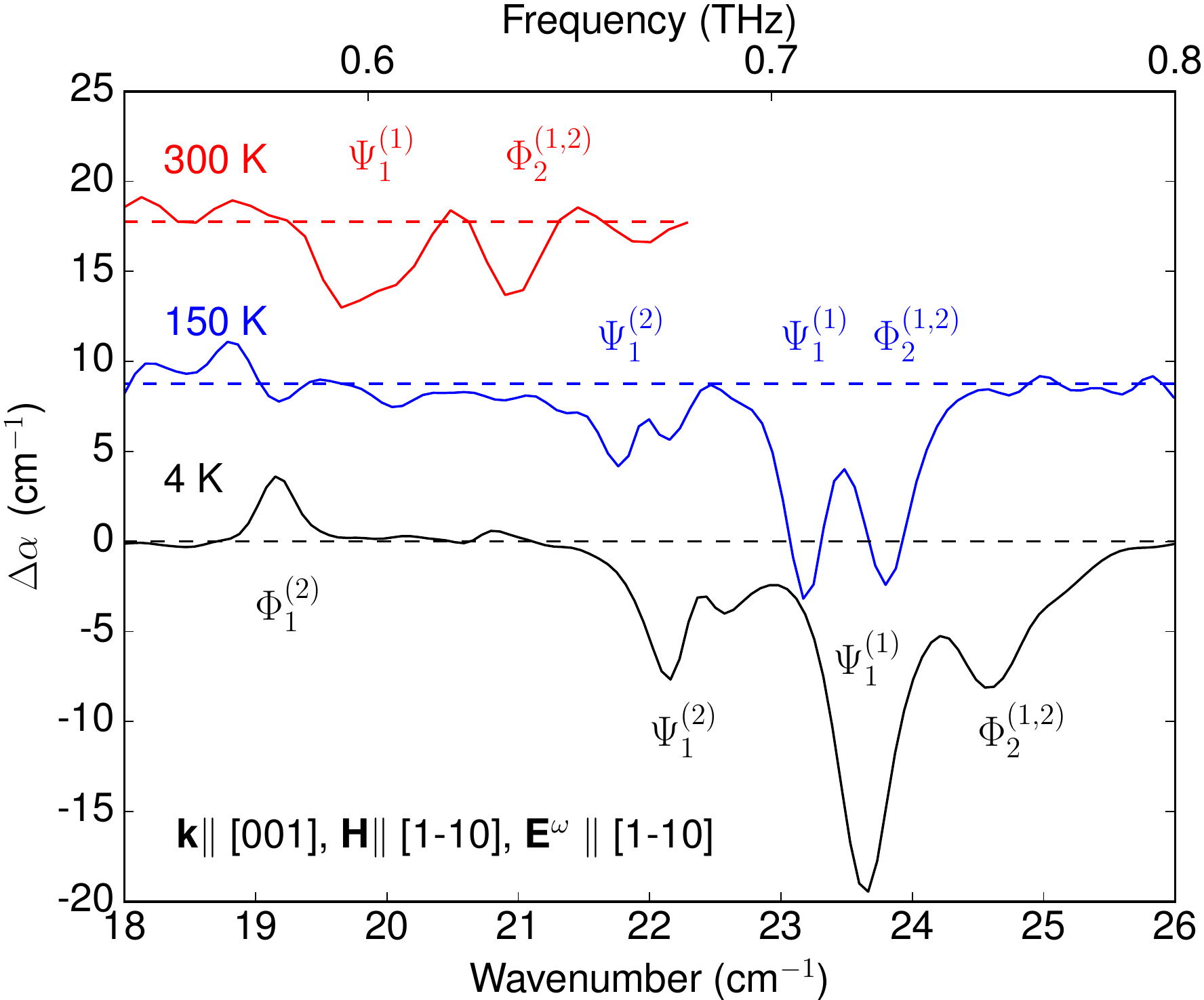}
\caption{\label{fig:Toroidal_HT} \textbf{$\mid$ Temperature
dependence of the directional dichroism in BiFeO$_3$.} Directional
dichroism spectra measured in $\mu_0H$$=$$\pm$12\,T at $T$$=$4, 150
and 300\,K. The magnetic field was applied along [1$\bar{1}$0] and
$\mathbf{E}^{\omega}$$\parallel$[1$\overline{1}0$]. Modes
$\Psi_1^{(1)}$ and $\Phi_2^{(1,2)}$ soften and get weaker with
increasing temperature, but are still clearly observable even at
300\,K. The $\Psi_1^{(2)}$ is visible until 150\,K while the
$\Phi_1^{(2)}$ cannot be detected reliably already at 150\,K.}
\end{figure}
%===========

Finally, we turn to the temperature dependence of the directional
dichroism presented in Fig.~\ref{fig:Toroidal_HT} for
$\mathbf{H}$$\parallel$$\mathbf{E}^{\omega}$$\parallel$[1$\overline{1}0$].
With increasing temperature the magnon modes
soften~\cite{talbayev11} and both the mean absorption and the
directional dichroism are reduced. Nevertheless, the modes
$\Psi_1^{(1)}$ and $\Phi_2^{(1,2)}$ still exhibit considerable
directional dichroism, $\Delta\alpha_{H}$$\approx$5\,cm$^{-1}$ at
room temperature. At low temperatures, almost perfect unidirectional
transmission was observed for the lowest-energy mode $\Psi_0$ with
orthogonal light polarization
($\mathbf{E}^{\omega}$$\parallel$[110]). Though we expect the same
at room temperature, $\Psi_0$ is out of our limited spectral window
at high temperatures.

The emergence of strong optical ME effect and the corresponding
unidirectional transmission require the simultaneous breaking of the
space- and time-inversion symmetries by the coexistence of
ferroelectricity and (anti)ferromagnetism. While these optical ME
phenomena have been investigated recently in various materials
hosting multiferroicity at low temperatures, here we studied the
unidirectional transmission in the spin excitation spectrum of
BiFeO$_3$~\cite{Martin2010,Sando2013,Henron2014,Matsukura2015}, the
unique multiferroic compound offering a real potential for room
temperature applications up to date. We found that the optical ME
effect in BiFeO$_3$ is robust enough to generate considerable
directional dichroism in the gigahertz-terahertz range even at room
temperature. Based on the current progress achieved in the electric
control of the magnetization in BiFeO$_3$, we expect that the
magnetic switching of the transmission direction, demonstrated here,
can be complemented by the electric control of the optical ME
effect. Because these functionalities exist at room temperature,
they can pave the way for the development of optical diodes with
electric and/or magnetic control.

\section*{Methods}

\textbf{Absorption measurements in the terahertz frequency range.}
The terahertz spectroscopy system consists of a Martin-Puplett type
interferometer with a Si bolometer operating at 300\,mK and a
mercury lamp. At high temperatures the spectral window of the
measurement was limited by the strong radiation load on the
detector. The light is directed to the sample using light pipes. The
sample is located in the He exchange gas filled sample chamber,
which is placed into the cold bore of a 17\,T superconducting
solenoid.

The measurement sequence was started by applying high magnetic
fields ($\geq$12\,T) at 4\,K for tens of minutes. For
$\mathbf{H}\parallel$ [110] and [1$\overline{1}$0], this procedure
respectively populates a single magnetic domain with $\mathbf{q}_1$
and two domains with $\mathbf{q}_2$ and $\mathbf{q}_3$. Next,
spectra were measured in different $\pm H$ fields. We did not see
any change in the magnetic domain population when the $-17$\,T field
was applied after $+17$\,T.

The zero field absorption spectrum was subtracted from the spectra
measured in finite fields. This procedure cancels out diffraction
and interference effects caused by the sample. The differential
absorption coefficient $\alpha(H)-\alpha(0)=-\ln[I(H)/I(0)]d^{-1}$,
where $I(0)$ and $I(H)$ are light intensity spectra in zero and $H$
field. The lower envelope of the whole set of differential
absorption spectra measured in different fields was used to
calculate the zero field spectrum. Magnetic field dependent
absorbtion spectra were evaluated as a sum of the zero field
spectrum and the corresponding differential spectra. Note that by
this method only the field-dependent part of this absorption is
recovered and field-independent features are not captured. While
this can cause an ambiguity of the mean absorption, the directional
dichroism spectrum $\Delta\alpha_H=\alpha(+H)-\alpha(-H)$ are free
of such uncertainties.

\textbf{Theoretical calculations.} The cycloid of BiFeO$_3$ is
controlled by two Dzyaloshinskii-Moriya (DM) interactions and an
easy-axis anisotropy $K$ along the ferroelectric polarization
$\vP_0$. Whereas the DM interaction $D_1$ perpendicular to $\vP_0$
is responsible for the formation of the long-period (62\,nm)
cycloids~\cite{pyatakov09,ed05}, the DM interaction $D_2$ along
$\vP_0$ is responsible for a small cycloidal tilt~\cite{pyatakov09}.

In a magnetic field $\vH$, the spin state and the magnon excitations
of \BF were evaluated from the microscopic Hamiltonian
\begin{eqnarray}
&&{\cal H} = -J_1\sum_{\langle i,j\rangle }\vS_i\cdot \vS_j
-J_2\sum_{\langle i,j \rangle'} \vS_i\cdot \vS_j
\nonumber \\
&&+D_1\, \sum_{\langle i,j\rangle} (\zp \times {\bf e}_{i,j}) \cdot (\vS_i\times\vS_j) \nonumber \\
&& + D_2\, \sum_{\langle i,j\rangle} \, (-1)^\ni \,\zp \cdot  (\vS_i\times\vS_j)\nonumber \\
&& -K\sum_i (\zp \cdot \vS_i )^2 - 2\mb \mathbf{H} \cdot\sum_i \vS_i
, \label{Ham}
\end{eqnarray}
$J_1$ and $J_2$ are the nearest and the second nearest neighbour
interactions, respectively. The $D_1$ sum, first proposed by Katsura
and coworkers~\cite{katsura05}, is uniform over the lattice, while
the $D_2$ sum alternates sign from one hexagonal layer to the next.

The nearest- and next-nearest neighbor exchange interactions
$J_1$$=$-5.32\,meV and $J_2$$=$-0.24\,meV were obtained from recent
inelastic neutron scattering measurements~\cite{matsuda12}.
$D_1$$\approx$0.18\,meV is determined from the cycloidal pitch.
$D_2$$=$0.085\,meV and $K$$=$0.0051\,meV were obtained by fitting
the four magnon modes observed in zero field~\cite{talbayev11,
nagel13}. The spin state of \BF is solved by using a trial spin
state that contains harmonics of the fundamental ordering wavevector
$\vQ $. We then minimize the energy $\langle {\cal H}\rangle $ over
the variational parameters of that state. A $1/S$ expansion about
the classical limit is used to evaluate the magnon mode frequencies
at $\vQ $ as a function of field~\cite{fishman13a, fishman13b}. We
calculated the optical ME susceptibilities,
$\hat{\chi}^{me}(\omega)$ and $\hat{\chi}^{em}(\omega)$, the
dielectric permittivity, $\hat{\varepsilon}(\omega)$, and the
magnetic permeability, $\hat{\mu}(\omega)$, using the Kubo
formula~\cite{Kezsmarki2014,Miyahara2012}.

To determine the directional dichroism, the Maxwell equations were
numerically solved for linearly polarized monochromatic plane waves
with $\pm\mathbf{k}$. Polarization rotation of the transmitted beam
was found negligible. We also evaluated the directional dichroism
spectra using the approximate formula in Eq.~(1), which is valid
when the polarization rotation of light can be
neglected~\cite{Kezsmarki2014}. Eq.~(1) and the numerical solution
of the Maxwell equations provide nearly equivalent
$\Delta\alpha_{k}$ spectra if parameters $\hat{\chi}^{me}(\omega)$,
$\hat{\chi}^{em}(\omega)$, $\hat{\varepsilon}(\omega)$ and
$\hat{\mu}(\omega)$ realistic to BiFeO$_3$ are used. Fits to the
directional dichroism spectra indicate that the polarization induced
by the spin current associated with $D_1$ and $D_2$ can be written
in the form given by Eq.(2), where $\lambda^{(1)}$ and
$\lambda^{(2)}$ are the dynamic ME couplings for the two types of
terms, respectively.

\textbf{Acknowledgements} We thank D. Szaller, S. Miyahara, N.
Furukawa and K. Penc for useful discussions. This work was supported
by the Estonian Ministry of Education and Research  Grant IUT23-03
and by the Estonian Science Foundation Grant ETF8703; by the
Hungarian Research Funds OTKA K 108918, OTKA PD 111756 and Bolyai
00565/14/11. The research at Oak Ridge National Laboratory was
sponsored by the Department of Energy, Office of Sciences, Basic
Energy Sciences, Materials Sciences and Engineering Division. The
work at Rutgers University was supported by the DOE under Grant No.
DOE: DE-FG02-07ER46382.

\textbf{Author Contributions} U.N. and T.R. performed the
measurements; U.N., T.R., S.B. analysed the data; S-W.C. and H.T.Y.
contributed to the sample preparation; R.S.F. and J.H.L. developed
the theory; I.K. wrote the manuscript; U.N., T.R. and I.K. planned
the project.

\textbf{Additional information} The authors declare no competing
financial interests.

\end{document}